\documentclass[aps,prc,reprint,a4paper,showpacs,showkeys,amsfonts,amssymb,amsmath,nofootinbib,twocolumn]{revtex4-1}
\usepackage{bm,amsmath,latexsym}
\usepackage{graphicx}

\usepackage[english]{babel}
\usepackage{hyperref}
\hypersetup{colorlinks=true}
\hypersetup{urlcolor=black}
\hypersetup{linkcolor=blue}
\hypersetup{citecolor=blue}
\usepackage[T1]{fontenc}
\usepackage{txfonts}

\newcommand{\dd}{\mathrm{d}}

\newcommand{\vp}{{\mathbf{p}}}

\newcommand{\bg}{\begin{align}}
\newcommand{\eeg}{\end{align}}
\newcommand{\be}{\begin{equation}}
\newcommand{\ee}{\end{equation}}
\newcommand{\ba}{\begin{eqnarray}}
\newcommand{\ea}{\end{eqnarray}}

\newcommand{\nn}{\nonumber}

\newcommand{\la}{\langle}
\newcommand{\ra}{\rangle}

\renewcommand{\geq}{\geqslant}
\renewcommand{\leq}{\leqslant}
\newcommand{\rspo}{\!\!\!\!\!\!\!\!}
\newcommand{\rsps}{\!\!\!\!\!\!}

\begin{document}

\title{Nucleon-Nucleon Interactions from Dispersion Relations: Coupled Partial Waves}
\author{M.~Albaladejo}
\email{albaladejo@um.es}
\author{J.~A.~Oller}
\email{oller@um.es}
\affiliation{Departamento de F\'{\i}sica. Universidad de Murcia, E-30071, Murcia, Spain.}

\begin{abstract}
We consider nucleon-nucleon  interactions from chiral effective field theory applying the $N/D$ method. The case of coupled partial waves is now treated, extending Ref.~\cite{paper1}, where the uncoupled case was studied. As a result, three $N/D$ elastic-like equations have to be solved for every set of three independent coupled partial waves.  As in the previous reference the input for this method is  the discontinuity along the left-hand cut of the nucleon-nucleon partial wave amplitudes. It can be calculated perturbatively in chiral perturbation theory because it involves only irreducible two-nucleon intermediate states. We apply here our method to the leading-order result consisting of one-pion exchange as the source for the discontinuity along the left-hand cut. The linear integral equations for the $N/D$ method must be solved in the presence of $\ell - 1$ constraints, with $\ell$ the orbital angular momentum, in order to satisfy the proper threshold behavior for $\ell\geq 2$. We dedicate special attention to satisfy the requirements of unitarity in coupled channels. We also focus on the specific issue of the deuteron pole position in the $^3S_1\text{--}^3D_1$ scattering. Our final amplitudes are  based on dispersion relations and chiral effective field theory, involving only convergent integrals. They are amenable to a systematic improvement order by order in the chiral expansion.
\end{abstract}

\maketitle
\section{Introduction}\label{sec:intro}

Recently we employed, in Ref.~\cite{paper1}, the $N/D$ method \cite{chew} to study nucleon-nucleon ($NN$) uncoupled partial-waves in connection with chiral perturbation theory (ChPT) \cite{physica,gl}. In this approach the chiral counting is applied to the imaginary part of the $NN$ partial waves along the left-hand cut (LHC),  owing to (multi-)pion exchanges. This can be done because  Cutkosky's rules require putting pion lines on-shell to calculate the discontinuity across the LHC, giving rise to irreducible nucleon diagrams. For more details see  Ref.~\cite{long}. At this point we avoid calculating perturbatively contributions that involve $N$-nucleon reducible graphs.\footnote{The latter require an extended version of the standard chiral counting as derived in Ref.~\cite{long}.} This method provides $NN$ partial waves by solving a linear integral equation that, by construction, involves convergent integrals and subtraction constants that can be calculated in terms of physical quantities. As a result, no need for any type of cutoff arises in our novel approach. 

The idea of applying  ChPT to evaluate irreducible $N$-nucleon contributions was originally put forward in Refs.~\cite{weinn}. There it was applied to calculate the effective multi-nucleon potential, which is later implemented in a Lippmann-Schwinger (LS) equation (or Schr\"odinger equation)  in order to derive the full $S$-matrix. However, owing to the singular nature of the chiral potentials resulting from their calculation in ChPT the solution of the LS equation  requires of some kind of regularization, typically a three-momentum cutoff $\Lambda$ \cite{ordo94,entem,epen3lo}. Several works \cite{nogga,pavon06,pavon06b,entem08,kswa} have shown that the chiral counterterms that appear in the ChPT potential following the standard ChPT counting \cite{weinn} are not enough to reabsorb the cutoff dependence that stems from the solution of the LS equation. Stable results with the $NN$ potential determined from one-pion exchange (OPE) are obtained in Refs.~\cite{nogga,phillipssw} for $\Lambda \rightarrow{\infty}$, by promoting  counterterms from higher orders to lower ones.\footnote{In Ref.~\cite{nogga} the cutoff range was taken $\Lambda<4$~GeV.} This implies a violation of the standard ChPT counting and of  the low-energy theorems relating the parameters in the effective range expansion \cite{eper}. One counterterm is needed for each partial wave with an attractive OPE tensor force \cite{nogga}, so that the $NN$ scattering amplitude from the OPE potential would require an infinity number of them, so that it is nonrenormalizable. A result compatible with this conclusion is also obtained in Ref.~\cite{eiras:2003}, where it is found that the $NN$ scattering amplitude from the OPE potential is nonrenormalizable unless the tensor force part vanishes for  $\Lambda\to \infty$ \cite{eiras:2003}. One should be aware that when $\Lambda \to \infty$ a more involved counting emerges \cite{birse,pavones,eiras:2003}. The extension of these ideas to higher orders in the chiral potential is not straightforward and, up to now, cannot avoid cutoff dependence \cite{phillipssw,pavonnew}. On the other hand, the application of Weinberg's scheme has given rise to a great phenomenological success in the reproduction of $NN$ phase shifts if the cutoff is fine-tuned in a region around 600~MeV, not  beyond the breakdown scale of the effective field theory (EFT) \cite{entem,epen3lo}. Of course, the cutoff dependence is not removed then.

We present the generalization of Ref.~\cite{paper1} to the case of coupled channels in Sec.~\ref{sec:formalism}, where the corresponding three linear integral equations needed for each set of coupled partial waves are derived. We apply this method to leading order (LO), which implies taking OPE as the source for the discontinuity along the LHC. References \cite{wong1, scotti63, scotti, oteo:1989} applied the $N/D$ method to study $NN$ scattering quantitatively. Reference \cite{wong1} was restricted to the $S$ waves and took only OPE as input along the LHC. References \cite{scotti63,scotti} included other heavier mesons as the source for the discontinuity along the LHC, in line with the meson theory of nuclear forces, so popular those days, while Ref.~\cite{oteo:1989} modeled the LHC discontinuity by OPE and one or two ad-hoc poles. We stress that we present here a novel way to introduce the $N/D$ method in harmony with the modern perspective of EFT. In this way, we show that one can calculate systematically within ChPT, according to the standard chiral counting, the discontinuity along the LHC that is the basic input for the $N/D$ method. This allows one to improve the results order by order, which was not the case by applying previous schemes \cite{Bjorken:1960zz, Bjorken2, noyes, oteo:1989}. This is a point of foremost importance. 

In addition, the threshold behavior of partial waves with orbital angular momentum $\ell \geqslant 2$ is satisfied within our approach by including zeroes at $\infty$ in the $NN$ partial waves, which is always allowed within the $N/D$ method (these are the so called Castillejo-Dalitz-Dyson (CDD) poles \cite{cdd}). However, in the previous works \cite{scotti63,scotti} the correct threshold behavior was achieved in an ad-hoc way by including a fictitious  pole below threshold, with the subsequent dependence of the results on its location, which was fitted to data. Furthermore, in our results we always respect coupled-channel unitarity, which was not the case in Refs.~\cite{wong1,scotti63,scotti,oteo:1989}.

The results obtained with this formalism are first considered for $^3S_1\text{--}^3D_1$ coupled waves in Sec.~\ref{sec:deuteron}, where the specific issue of the deuteron pole is also discussed. Higher partial waves are considered in  Sec.~\ref{sec:results}. Our conclusions are collected in Sec.~\ref{sec:conclusions}. Finally, we show in the Appendix the cancellation of a potential divergence in a function involved in our equations. This cancellation occurs thanks to the constraints already imposed to satisfy the right threshold behavior for partial waves with $\ell\geq 2$.

\section{Coupled partial waves}
\label{sec:formalism}
For spin triplet $NN$ partial waves with total angular momentum $J$ one has the mixing of the orbital angular momenta $\ell = J-1$ and $\ell' =J+1$   (except for the $^3P_0$ partial wave). Each set of coupled partial waves is determined by the quantum numbers $S$, $J$, $ \ell$, and $\ell'$, where $S$ is the total spin. In the following, to simplify the notation we omit them and indicate the different partial waves by $t_{ij}$, with $i=1$ corresponding to $\ell=J-1$ and $i=2$ to $\ell'=J+1$, a convention that we adopt henceforth. As a result, a two-coupled-channel $T$-matrix results. In our normalization, the resulting $S$-matrix reads
\begin{align}
S(A) & = \mathbb{I} + i 2 \rho(A) T(A) = \nn\\
& = \left(
 \begin{array}{cc}
 \cos 2\epsilon\ e^{2i\delta_1}            & i\sin 2\epsilon\ e^{i(\delta_1+\delta_2)} \\ 
i\sin 2\epsilon\ e^{i(\delta_1+\delta_2)} &   \cos 2\epsilon\ e^{2i\delta_2}
 \end{array} \right)~\text{,}
\label{relst}
\end{align}
 such that $\delta_1$ corresponds to the phase shifts for the channel with $\ell = J-1$ and $\delta_2$ to that with $\ell' = J+1$. The argument $A$ refers to the center-of-mass (CM) three-momentum squared. We have indicated the phase space by $\rho(A)$, which reads, in our non-relativistic approximation, $\rho(A) = m\sqrt{A}/4\pi$, where $m$ is the nucleon mass.

Along the right-hand cut (RHC), which corresponds to the physical region with $A>0$, the unitarity of the $S$-matrix, $S S^\dagger = S^\dagger S = \mathbb{I}$~, can be written in terms of the (symmetric) $T$-matrix as $\mathrm{Im} T^{-1}(A) = - \rho(A)\, \mathbb{I}$. In the following, the imaginary parts above threshold of the inverse of the $T$-matrix elements, denoted $t_{ij}(A)$,  play an important role, 
\begin{align}
\mathrm{Im} \frac{1}{t_{ij}(A)} \equiv -\nu_{ij}(A)~,A>0~\text{.}
\label{nuij.def}
\end{align}
 Employing the relationship between the $T$- and $S$-matrices, Eq.~\eqref{relst}, 
 we can express the different $\nu_{ij}$ in terms of phase shifts and the mixing angle along the physical region above threshold. In this way, one can write the diagonal partial waves as $t_{ii} = (e^{2i\delta_i} \cos 2\epsilon -1)/2i\rho$, while for the mixing amplitude $t_{12}=e^{i(\delta_1+\delta_2)}\sin2\epsilon/2\rho$. From these expressions it is straightforward to obtain, for $A>0$: 
\begin{align}
\nu_{11}(A) & =   \rho(A) \left[ 1- \frac{\frac{1}{2}\sin^2 2\epsilon}{1-\cos 2\epsilon \cos 2\delta_1} \right]^{-1} \label{eq:nus11}~,\\
\nu_{22}(A) & =   \rho(A) \left[ 1- \frac{\frac{1}{2}\sin^2 2\epsilon}{1-\cos 2\epsilon \cos 2\delta_2} \right]^{-1} \label{eq:nus22}~,\\
\nu_{12}(A) & = 2 \rho(A) \frac{\sin(\delta_1 + \delta_2)}{\sin 2\epsilon} \label{eq:nus12}~.
\end{align}
Although not explicitly indicated, it should be understood that the phase shifts and  mixing angle depend on $A$. Equation \eqref{nuij.def} generalizes that of an uncoupled partial wave, $\mathrm{Im} T^{-1} = -\rho$, employed in Ref.~\cite{paper1}. Indeed, if we set $\epsilon = 0$ in $\nu_{11}(A)$ and $\nu_{22}(A)$, the uncoupled case is recovered. Note also that $\nu_{ii}(A)/\rho(A) \geq 1$.

\begin{figure}[t]
\includegraphics[width=5cm,keepaspectratio]{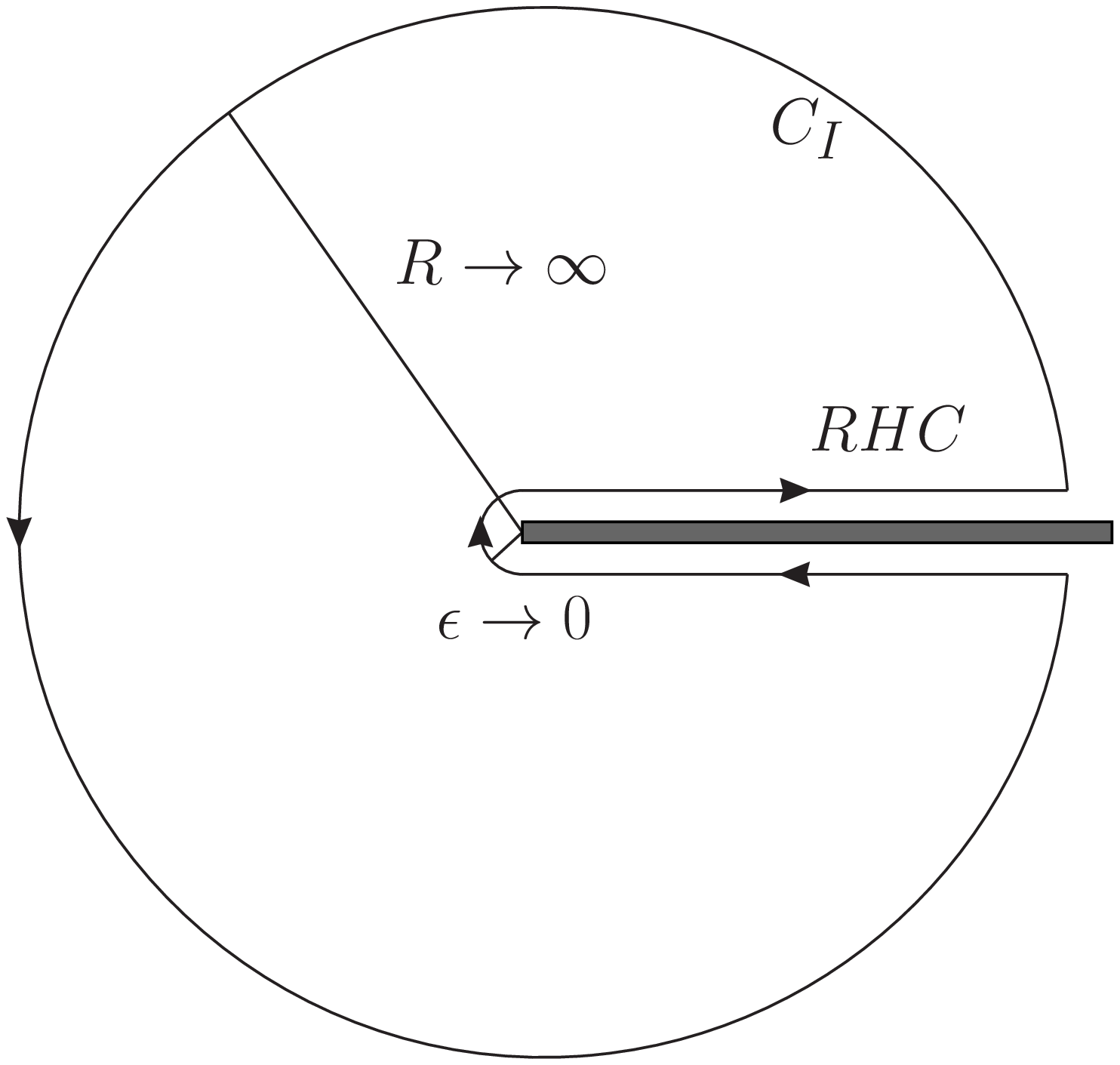}\\
\includegraphics[width=5cm,keepaspectratio]{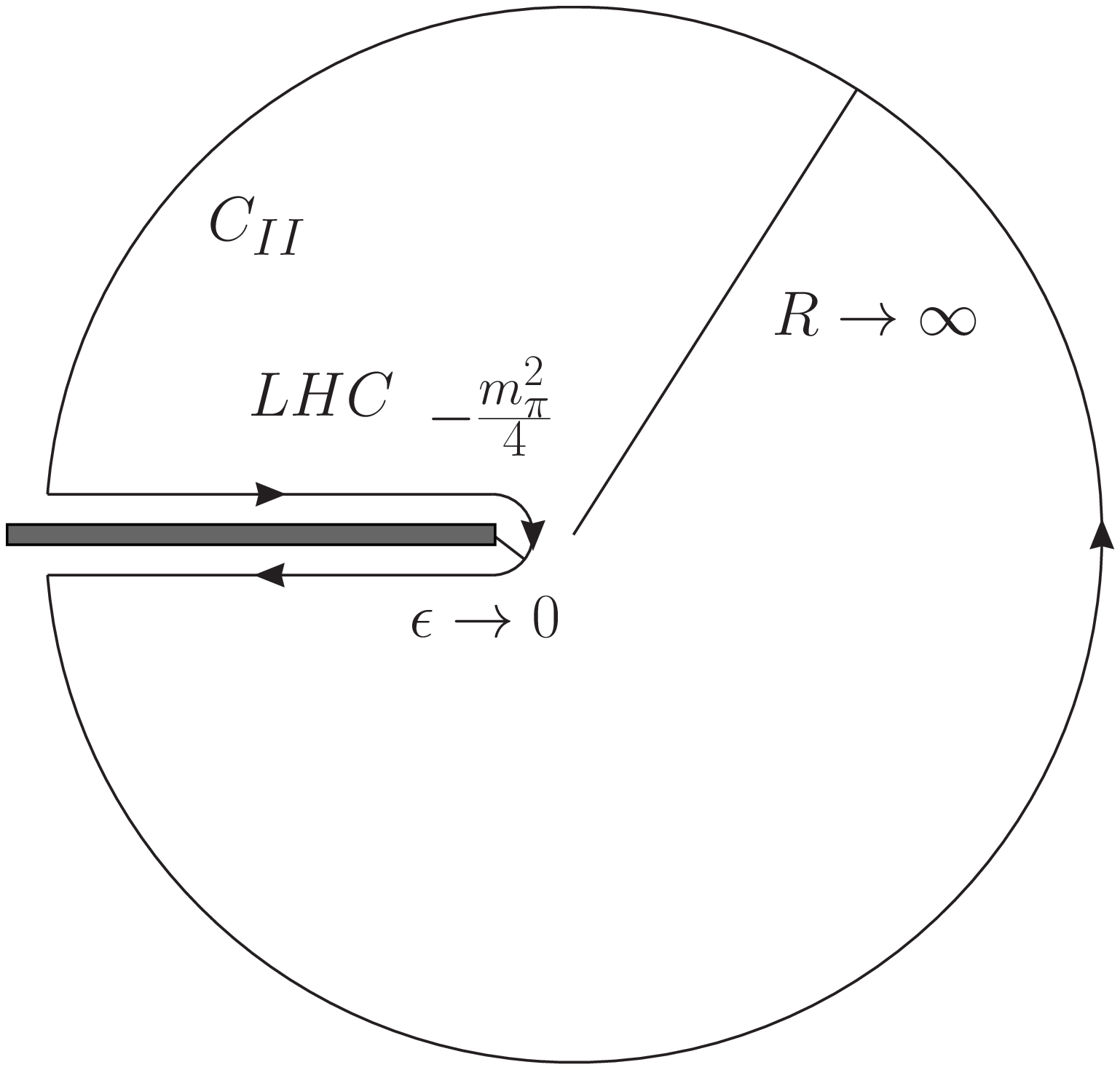}
\caption{The thick lines correspond to the RHC and LHC, from top to bottom. In the same figure the integration contours $C_I$ and $C_{II}$ for evaluating $D_{ij}(A)$ and $N_{ij}(A)$, respectively, are shown. One has to take the limit $\epsilon\to 0^+$.
\label{fig:cut}}
\end{figure}

We apply the $N/D$ method \cite{chew} to solve our equations for the $T$-matrix. A general $NN$ partial wave has two types of cuts, the LHC and RHC, the former due to crossed channel dynamics and the latter to unitarity. The lightest particle that is exchanged between two nucleons is the pion, which determines the onset of the LHC for $A < L\equiv -m_\pi^2/4$, with $m_\pi$ the pion mass. The unitarity cut occurs for $A>0$. See Fig.~\ref{fig:cut}, where the LHC and RHC are indicated separately. In the $N/D$ method a partial wave $t_{ij}$ is written as the quotient of a numerator function $N_{ij}(A)$ and a denominator one $D_{ij}(A)$. The function $N_{ij}$ only has a LHC while the function $D_{ij}$ has only a RHC. In Refs.~\cite{Bjorken:1960zz,Bjorken2}, a straightforward generalization of the one-channel $N/D$ method of Chew and Mandelstam \cite{chew} was given by writing $T=N \cdot D^{-1}$ in matrix notation. This $T$-matrix would be symmetric, as it is required by temporal inversion, only under the assumption that $D^T(T^T-T)D$ vanishes for $A\to \infty$ \cite{Bjorken2}, where the superscript $T$ indicates the transpose of the corresponding matrix. However, this is not the case for the chiral potentials, even at LO, e.g., in the $^3S_1\text{--}^3D_1$ coupled partial waves. This condition is thus too restrictive for its application to chiral EFT, where different numbers of subtractions are taken in the different partial waves involved, whose number also varies according to the chiral order considered in the calculation of the imaginary part of the $NN$ partial wave amplitude along the LHC.

In what follows, we generalize the procedure of Ref.~\cite{paper1} to the coupled case. Instead of making use of a matrix notation as in Refs.~\cite{Bjorken:1960zz,Bjorken2}, we write three $N/D$ equations, one for each of the three independent partial waves $t_{ij}$, as in Ref.~\cite{noyes},
\begin{equation}
\label{nd.def.cou}
t_{ij}(A) = A^{\ell_{ij}} \frac{N_{ij}(A)}{D_{ij}(A)}~.
\end{equation}
The factor $A^{\ell_{ij}}$ guarantees the proper  threshold behavior with  $\ell_{11} = \ell$, $\ell_{22} = \ell' = \ell + 2$ and $\ell_{12} = (\ell + \ell')/2 = \ell + 1$. We focus here on the specific features of the coupled channel mechanism, referring the reader to Ref.~\cite{paper1} for further details on the general procedure followed to apply an $N/D$ equation. As stated above the splitting of the $t_{ij}(A)$ function is such that $N_{ij}$ bears the LHC and $D_{ij}$ the RHC, and then:
 \begin{align}
 \text{Im}D_{ij}(A)&=-N_{ij}(A) A^{\ell_{ij}} \nu_{ij}(A)~,~A>0~, \\
 \text{Im}N_{ij}(A)&= D_{ij}(A) \Delta_{ij}(A)/A^{\ell_{ij}}~,~A < L ~,
 \end{align}
with $\mathrm{Im}t_{ij} \equiv \Delta_{ij}$ along the LHC. The imaginary parts of $D_{ij}$ and $N_{ij}$ are $0$ elsewhere along the $A$-real axis.\footnote{Because the Schwartz reflection principle is satisfied by $t_{ij}$, $D_{ij}$ and $N_{ij}$ the discontinuity across the RHC or LHC is given by $2i$ times the imaginary part of the function.} As argued in Refs.~\cite{long, paper1}, $\Delta_{ij}$ can be calculated perturbatively in   
 ChPT along the LHC, as it originates from multi-pion exchanges putting pion propagators on-shell. The intermediate states thus require at least one pion so that we apply ChPT always to irreducible $N$-nucleon diagrams, responsible for the discontinuity along the LHC. 
 
Two dispersion relations (DRs) can be written for the functions $D_{ij}$ and $N_{ij}$, employing the contours $C_I$ and $C_{II}$ in Fig.~\ref{fig:cut}, respectively. The integration along the circle at infinity vanishes, if necessary, by taking sufficient number of subtractions. At LO in the chiral counting \cite{let1,long}, the only contribution to $\Delta_{ij}$ along the LHC is OPE. Asymptotically, for $\vp^2\to -\infty$, OPE tends to constant, so that, according to the Sugawara and Kanazawa theorem \cite{barton,suga} one subtraction is 
necessary for the DR of $N_{ij}(A)$ in $S$ wave, even though $\Delta_{ij}(A)\to 1/A$ in the case of OPE.
 On general grounds, a partial wave amplitude is bounded because of unitarity by constant/$\sqrt{A}$ for $A\to +\infty$ so that $t_{ij}D_{ij}(A)/A^{\ell_{ij}}$ tends to constant for an $S$ wave and $0$ for any other partial wave.\footnote{Here we are taking that $D_{ij}$ diverges as $\sqrt{A}$ for $A\to \infty$ as in the uncoupled case \cite{paper1}. This is consistent with the results obtained explicitly in this work.} As a result the same theorem then requires that at least one subtraction is necessary for the $S$ waves:
 \begin{align}
 \label{dd.cou}
D_{ij}(A)&=1-\frac{A}{\pi}\int_0^{+\infty}\rspo  \dd q^2\frac{\nu_{ij}(q^2)N_{ij}(q^2)q^{2\ell_{ij}}}{q^2(q^2-A)}
~,\\
\label{nn.cou.3s1}
N_{ij}(A)&=N_0 + \frac{A}{\pi}\int_{-\infty}^{L}\rsps  \dd k^2
\frac{\Delta_{ij}(k^2)D_{ij}(k^2)}{k^{2}(k^2-A)}~~,~~\ell_{ij} = 0~,\\
\label{nn.cou.gen}
N_{ij}(A)&=\frac{1}{\pi}\int_{-\infty}^{L}\rsps \dd k^2
\frac{\Delta_{ij}(k^2)D_{ij}(k^2)}{k^{2\ell_{ij}}(k^2-A)}~~,~~\ell_{ij}\neq 0
~.
\end{align}
The subtraction point is taken at threshold (see Ref.~\cite{paper1} for expressions with the subtraction point at any other position). One subtraction is taken for the $D_{ij}(A)$ function, which is fixed to 1 because, in view of Eq.~\eqref{nd.def.cou}, only the ratio $N_{ij}/D_{ij}$ matters in order to determine $t_{ij}$. Thus, there is the freedom to fix the value of $D_{ij}$ at one point, e.g., at threshold, by simultaneously dividing $D_{ij}$ and $N_{ij}$ by the appropriate constant. For $\ell_{ij}=0$, $S$ wave, one subtraction is taken in $N_{ij}(A)$, as just discussed. In our present work, dedicated to the $NN$ coupled partial waves, this is the case only for the $^3 S_1$ channel. The subtraction constant $N_0$ is the amplitude at threshold, $t_{11}(0) = N_0$, and then it can be fixed in terms of the $^3S_1$ scattering length, $a_t$,
\begin{equation}
N_0 = - \frac{4\pi a_t}{m}~,
\label{n0.fix}
\end{equation}
with the value $a_t = 5.424 \pm 0.004\ \text{fm}$. Below in Sec.~\ref{sec:deuteron} we also fix $N_0$ in terms of the experimental deuteron binding energy.

An integral equation for the function $D_{ij}(A)$ results by inserting Eqs.~\eqref{nn.cou.3s1} or \eqref{nn.cou.gen} into Eq.~\eqref{dd.cou}. However, as argued in detail in Ref.~\cite{paper1}, divergent integrals appear for $\ell\geq 2$ unless a set of $\ell-1$ constraints is satisfied by $D_{ij}(A)$. These constraints are a generalization of those satisfied by OPE. We just quote the final result from Ref.~\cite{paper1}, which is given in terms of the set of sum rules:
\begin{align}
\int_{-\infty}^L \rsps \dd k^2 \frac{\Delta_{ij}(k^2)D_{ij}(k^2)}{k^{2\lambda}}=0~,\quad \lambda=2,3,\ldots,\ell_{ij} \geq 2
\label{n.sr.cou}
\end{align}
Expanding the denominator inside the integral of Eq.~\eqref{nn.cou.gen}, $N_{ij}(A)$ can be written as
\begin{align}
\label{nij.vanish}
N_{ij}(A)& = \frac{1}{\pi A^{\ell_{ij}-1}}\int_{-\infty}^L \rsps \dd k^2 \frac{\Delta_{ij}(k^2)D_{ij}(k^2)}{k^2(k^2-A)} \nn \\
&-\frac{1}{\pi}\sum_{m=0}^{\ell_{ij}-2}\frac{1}{A^{m+1}}\int_{-\infty}^L \rsps
 \dd k^2 \frac{\Delta_{ij}(k^2)D_{ij}(k^2)}{k^{2(\ell_{ij}-m)}}~,
\end{align}
and the terms within the sum vanish if the constraints of Eq.~\eqref{n.sr.cou} are fulfilled. This guarantees that $N_{ij}(A)$ vanishes as $ 1/A^{\ell_{ij}}$, which ensures the convergence of the resulting integral equation for $D_{ij}(A)$. 

Let us take first $\ell_{ij}\neq 0$. By inserting the nonvanishing piece of $N_{ij}$ into Eq.~\eqref{dd.cou}, once the constraints Eq.~\eqref{n.sr.cou} are satisfied, we find the following integral equation for $D_{ij}(A)$:
\begin{align}
\label{int.d.nodiv.cou}
D_{ij}(A)& = 1+\frac{A}{\pi}\int_{-\infty}^L \rsps \dd k^2 \frac{\Delta_{ij}(k^2)D_{ij}(k^2)}{k^2} g_{ij}(A,k^2)~,\\
\label{gij_fun_def}
g_{ij}(A,k^2)& = \frac{1}{\pi}\int_0^{+\infty} \rspo \dd q^2 \frac{\nu_{ij}(q^2)}{(q^2-A)(q^2-k^2)}~.
\end{align}
The functions $g_{ij}(A,k^2)$ are the generalization of $g(A,k^2)$ given in Ref.~\cite{paper1} for the uncoupled case. An important technical detail is discussed in the Appendix. We show there how the constraints in Eq.~\eqref{n.sr.cou} guarantee that the  functions $g_{ij}(A,k^2)$ are finite curing a potential divergence for $ij=22$ in the $q^2\to 0$ limit. This divergence was noticed in Ref.~\cite{noyes} but no procedure for removing it was given there.

The $N/D$ method in the presence of the constraints, Eq.~\eqref{n.sr.cou}, was solved  in Ref.~\cite{paper1} by means of the insertion of CDD poles \cite{cdd} by taking advantage of the fact that  the $D_{ij}$ functions are determined modulo the addition of CDD poles \cite{barton,spearman,mandelstam}. The main points from Ref.~\cite{paper1}, briefly summarized, consist of using this ambiguity  to include $\ell_{ij}-1$ CDD poles (if $\ell_{ij} \geq 2$) in $D_{ij}(A)$. These poles are gathered at the same position $B$, and finally the limit $B\to \infty$ is taken. The following equations are then obtained \cite{paper1}:
\begin{align}
\label{nn.final.cou}
N_{ij}(A)& = \frac{1}{\pi}\int_{-\infty}^L \rsps \dd k^2 \frac{\Delta_{ij}(k^2)D_{ij}(k^2)}{k^{2\ell_{ij}}(k^2-A)}~,\\
\label{dd.final.cou}
D_{ij}(A)& = 1+\frac{A}{\pi}\int_{-\infty}^L \rsps \dd k^2 \frac{\Delta_{ij}(k^2)D_{ij}(k^2)}{k^2} g_{ij}(A,k^2) \nn \\
& + \frac{A \sum_{n=0}^{\ell_{ij}-2} c_n A^n}{(A-B)^{\ell_{ij}-1}}~.
\end{align}
The last sum corresponds to the addition of the $\ell_{ij}-1$ CDD poles. The  coefficients $c_i$ are determined in such a way that the constraints in Eq.~\eqref{n.sr.cou} are satisfied (see Ref.~\cite{paper1} for further details).

Note that for the  $P$ waves ($\ell_{ij}=1$) (in the present study we have the mixing partial wave in the $^3 S_1\text{--}^3 D_1$ system and  the $^3 P_2$ in $^3P_2\text{--}^3F_2$ scattering), no constraints are needed \cite{paper1}, so that the sum over the CDD poles is dropped and the same formalism applies. This is also clear because, for this case, Eq.~\eqref{nn.cou.gen} vanishes as $1/A$ so that there is no room for restrictions.

Let us take now the case $\ell_{ij}=0$, which only occurs for the $^3 S_1$ wave. Since a subtraction is needed in $N_{11}$, Eq.~\eqref{nn.cou.3s1}, one should change Eq.~\eqref{dd.final.cou} in two ways, as there is no sum over CDD poles and one has to include an extra term associated with the subtraction in $N_{ij}(A)$ for this case. It is straightforward to obtain, by inserting Eq.~\eqref{nn.cou.3s1} into Eq.~\eqref{dd.cou}, the appropriate integral equation for $D_{11}(A)$ for the $^3S_1$ partial wave:
\begin{align}
\label{dd.final.3s1}
D_{11}(A)& = 1 - A N_0 g_{11}(A,0) \nn \\
& + \frac{A}{\pi}\int_{-\infty}^L \rsps \dd k^2 \frac{\Delta_{11}(k^2)D_{11}(k^2)}{k^2} g_{11}(A,k^2)~,
\end{align}
with $g_{11}(A,k^2)$ given by Eq.~\eqref{gij_fun_def}. Notice also that, from Eq.~\eqref{nn.cou.3s1}, it is clear that $N_{ij}(A)$ tends to constant for $\ell_{ij}=0$ and $A\to \infty$, so that there is no need for constraints. This is why no sum over CDD poles is present in the previous equation. 

To obtain the final amplitudes, the $D_{ij}(A)$ functions are obtained along the LHC ($A<-m_\pi^2/4$) by solving the integral equations in Eq.~\eqref{dd.final.cou} or Eq.~\eqref{dd.final.3s1}. Next, the functions $D_{ij}(A)$ are obtained along the RHC ($A>0$) from the same equations  because the integrand is known. To obtain the functions $N_{ij}(A)$, since the constraints in Eq.~\eqref{n.sr.cou} are obeyed, one can use for $\ell_{ij}\neq 0$ either Eq.~\eqref{nn.final.cou} or the first term on the right-hand side of Eq.~\eqref{nij.vanish} (but the former is more suitable numerically, since it converges faster). For the $^3S_1$ wave, one should use Eq.~\eqref{nn.cou.3s1}. The partial waves $t_{ij}(A)$ are obtained by employing the resulting $D_{ij}(A)$ and $N_{ij}(A)$ functions in Eq.~\eqref{nd.def.cou}. 

The main difference with respect to the uncoupled case treated in Ref.~\cite{paper1} is that now one has to solve simultaneously three $N/D$ equations for $ij$=$11$, $12$ and $22$ with the functions $g_{ij}(A,k^2)$ linked between each other. They depend on the phase shifts $\delta_1$ and $\delta_2$ and on the mixing angle $\epsilon$, defined in Eq.~\eqref{relst}, which are also the final output of our approach. Thus, we employ an iterative procedure (similar to Ref.~\cite{noyes}) as follows. Given an input for $\delta_1$, $\delta_2$ and $\epsilon$, one solves the three integral equations for $D_{ij}(A)$ along the LHC, and then the amplitudes for the RHC can be calculated. The new phase shifts $\delta_1$ and $\delta_2$ are obtained from the phase of the $S$-matrix elements $S_{11}$ and $S_{22}$, while $\sin 2\epsilon=2\rho A^{\ell_{12}}N_{12}/\lvert D_{12} \rvert$, according to Eq.~\eqref{relst}. In this way a new input set of  functions $\nu_{ij}$, Eqs.~\eqref{eq:nus11}-\eqref{eq:nus12}, results.  These are used again in the integral equations, and the iterative procedure is finished when convergence is found (typically, the difference between one iteration in the three independent $D_{ij}$ functions along the LHC is required to be less than one per mil). As  initial input one can use the results given by UChPT \cite{long}, or some placed-by-hand phase shifts and mixing angle, and we find no dependence of our final unitary results on the input employed.

It can be shown straightforwardly that unitarity is fulfilled in our coupled channel equations, solved in the way just explained, if $|S_{11}(A)|^2=|S_{22}(A)|^2=\cos^2 2\epsilon$ for $A>0$. From the fact that $\hbox{Im}t_{12}=\nu_{12}|t_{12}|^2$, as follows from  Eq.~\eqref{nuij.def}, and $\sin2\epsilon=2\rho |t_{12}|$ (the latter equality is valid  only when convergence is reached), it follows that the phase of $t_{12}$ is $\delta_1+\delta_2$, as required by unitarity,  Eq.~\eqref{relst}. By construction the phase shifts are equal to one-half the phase of the $S$-matrix diagonal elements when convergence is achieved.

\section{Results}

We now present the reproduction of the phase shifts and mixing angles for the $NN$ coupled partial waves with $J\leq 3$ compared with the data from the Nijmegen partial wave analysis (PWA) \cite{Stoks:1994wp}. We pay special attention to the $^3S_1\text{--}^3D_1$ system.

\subsection{\boldmath $^3S_1\text{--}^3D_1$ coupled waves}
\label{sec:deuteron}

In this section we discuss our results for the $^3S_1\text{--}^3D_1$ coupled waves. Previous papers applying the $N/D$ method to adjust $NN$ scattering are Refs.~\cite{wong1,scotti63,scotti,oteo:1989}. We already commented about Ref.~\cite{wong1,oteo:1989}. The other two works, by Wong and Scotti \cite{scotti63,scotti}, include, together with OPE, other heavier mesons, $\eta$, $\rho$, $\omega$, and $\phi$ is also included in Ref.~\cite{scotti}. Thus, these works follow the basic ideas of meson theory of nuclear forces, that were also used for the construction of $NN$ potentials \cite{machleidt}. There are some approximations in Refs.~\cite{wong1,scotti63,scotti} that we avoid in our work. For example, only elastic unitarity is used in Refs.~\cite{wong1,scotti63} neglecting the mixing between coupled partial waves. Reference \cite{scotti} considers the mixing only for $^3S_1\text{--}^3D_1$ coupled partial waves. In addition, in order to satisfy the threshold behavior for partial waves with $\ell\geq 2$, so that they vanish as $A^\ell$, Refs.~\cite{scotti63,scotti} make use of a rather ad-hoc formula. This method was criticized in Ref.~\cite{noyes65} because it includes an unphysical pole for every partial wave at a CM squared energy $s_1$, somewhat below $4m^2$ (the threshold for $NN$ scattering). In addition, Refs.~\cite{scotti63,scotti} also have a cutoff dependence in the way the vector resonance exchanges damp to avoid their divergences at infinity. Though the results of Refs.~\cite{wong1,scotti63,scotti} are interesting and typically obtain a good reproduction of data at the phenomenological level, we offer here a novel way of employing the $N/D$ method in light of EFT. We then present the method ready to be used in a systematic way by improving, order by order, the discontinuity of the partial wave amplitudes along the LHC as it involves only $NN$ irreducible diagrams, as discussed above \cite{long}. We satisfy exact unitarity for all the partial waves as well. It is also important to stress that the $N/D$ method for coupled channels is now presented in a way ready to be used at any chiral order, without being constrained to satisfy the too demanding Bjorken-Nauenberg condition \cite{Bjorken2} in order to end with symmetric partial waves. We accomplish the right threshold behavior for $\ell\geq 2$ by adding CDD poles at infinity, which is always legitimate in the $N/D$ method if there are good reasons to include them (which have been offered before \cite{paper1}). Thus, we do not need to modify the right analytical properties of partial waves by including a fictitious pole in $s_1$ which is then fine tuned to data, as done in Refs.~\cite{scotti63,scotti}.  

The deuteron ($d$) is a neutron-proton ($np$) bound state with total angular momentum $J=1$ and spin $S=1$ (and isospin 0). As such, it is seen as a pole below threshold ($|\vp|^2 <0$) in the physical Riemann sheet in $^3S_1\text{--}^3D_1$ coupled partial waves. The binding energy of the deuteron, $E_d$ (defined positive), is given by
\begin{equation}
E_d = -\frac{k_d^2}{m}~,
\end{equation}
where $k_d^2$ is the three-momentum squared at which the pole is located, so that it is negative. Specifically, in our approach it appears as a zero in the functions $D_{ij}(A)$. From the amplitudes calculated in Sec.~\ref{sec:formalism} we find the deuteron at the position $k_d^2=-0.08 m_\pi^2$ in the $^3 S_1$ amplitude, corresponding to $E_d\simeq 1.7\ \text{MeV}$. Recall that the subtraction constant $N_0$ appearing in the $^3S_1$ partial wave is determined by fixing the $^3S_1$ scattering length to its experimental value, Eq.~\eqref{n0.fix}. There is still a remnant input dependence for the $^3S_1\text{--}^3D_1$ coupled partial waves in our unitary solutions that we fix by requiring that the deuteron pole position is the same in the $^3S_1$ and in the mixing partial wave. Independently of the input we do not find any pole in the $^3D_1$ partial wave. Indeed, if we disregard the coupling between $^3S_1$ and $^3D_1$ and use the method of Ref.~\cite{paper1} for uncoupled waves, the pole appears in the same position in $^3 S_1$ and, again, it does not appear in $^3 D_1$. However, the pole should be located at the same energy in every channel, but this is not the case because we are not using a matrix formalism but solving the three linked $N/D$ equations independently. Notice also that the deuteron is found mainly in a $^3 S_1$ state, and thus the coupling to $^3 D_1$ is very weak.

\begin{figure*}[t]
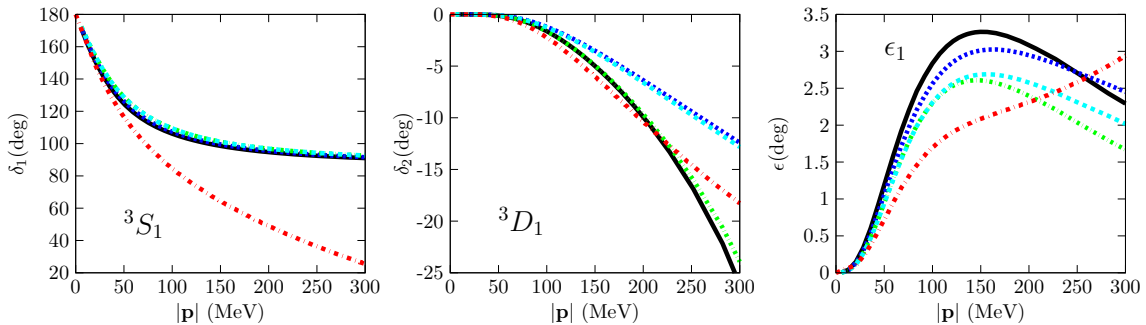
\centering
\begin{tabular}{ccc}
\includegraphics[height=4.2cm,keepaspectratio]{figurasNN_29.eps} &
\includegraphics[height=4.2cm,keepaspectratio]{figurasNN_30.eps} &
\includegraphics[height=4.2cm,keepaspectratio]{figurasNN_31.eps}
\end{tabular}
\caption{(Color online) Comparison of our results for the $^3 S_1$ and $^3 D_1$ phase shifts and the mixing angle $\epsilon_1$  to the Nijmegen PWA \cite{Stoks:1994wp} , shown by the dot-dashed (red) lines. The solid (black) lines correspond to fixing the $^3S_1$ scattering length to experiment, while the dashed (blue) lines in addition fix the deuteron pole position in $^3D_1$ at the same value as that in $^3S_1$. On the other hand, the double dotted (green) lines stem by fixing the deuteron pole position in the $^3S_1$ partial wave at its experimental value. The dash-double-dotted (cyan) lines correspond to having additionally fixed the deuteron pole in the $^3D_1$ partial wave at the same point as in $^3S_1$. 
\label{fig_3s13d1_deuteron}}
\end{figure*}

In order to cure this deficiency and having the right pole structure guaranteeing the presence of the deuteron pole in $^3D_1$, we write a twice subtracted DR for the $^3 D_1$ partial wave, such that the function $D_{22}(A)$ has a zero at a given $k_d^2$. The DR reads
\begin{align}
D_{ij}(A) & = 1 - \frac{A}{k_d^2} \nn \\
& - \frac{A(A-k_d^2)}{\pi} \int_0^{+\infty} \rspo \dd q^2~\frac{\nu_{ij}(q^2)N_{ij}(q^2)q^{2\ell_{ij}}}{q^2(q^2-A)(q^2 - k_d^2)}~,
\end{align}
written in a way that is valid both for the $^3D_1$ partial wave ($ij=22$) and for the mixing partial wave ($ij=12$), although we do not use it for the latter. By inserting the expression for $N_{ij}(A)$, Eq.~\eqref{nn.cou.gen}, into the previous equation, we end up with the following integral equation
\begin{align}
D_{ij}(A) & = 1 - \frac{A}{k_d^ 2} \nn\\
& +\frac{A(A-k_d^2)}{\pi}\int_{-\infty}^L \rsps \dd k^2 ~\frac{\Delta_{ij}(k^2)D_{ij}(k^2)}{k^{2\ell_{ij}}} g^{(d)}_{ij}(A,k^2)~,
\label{gdij}
\end{align}
where $g^{(d)}_{ij}(A,k^2)$ is a generalization of the functions $g_{ij}(A,k^2)$ in Eq.~\eqref{gij_fun_def},
\begin{equation}
g^{(d)}_{ij}(A,k^2) = \frac{1}{\pi} \int_0^{+\infty}  \rspo \dd q^2 ~ \frac{\nu_{ij}(q^2)
q^{2(\ell_{ij}-1)}}{(q^2-A)(q^2-k^2)(q^2 - k_d^2)}~.
\label{gijd}
\end{equation}
For $^3 S_1\text{--}^3 D_1$ waves, we have $\ell=0$ and $\ell'=2$, so that $\ell_{12} = 1$ and $\ell_{22}=2$, and the previous integrals are convergent because of the extra subtraction taken. Recall that, in the formalism first presented in Sec.~\ref{sec:formalism}, one must take into account a constraint for the $D_{22}(A)$ partial wave in order to end with a convergent integral equation. Note that from Eqs.~\eqref{gdij} and \eqref{gijd} the high-energy behavior of the functions $D_{ij}$ changes, now diverging as $A^{3/2}$, instead of $A^{1/2}$ as in Sec.~\ref{sec:formalism} or in Ref.~\cite{paper1}. As a result, the criterion of imposing that $N_{ij}\to 1/A^{\ell_{ij}}$ for $A\to \infty$, the one used in Ref.~\cite{paper1} to deduce the need of constraints, does not hold in this case because of the extra subtraction.\footnote{From Eq.~\eqref{nn.cou.gen} it follows immediately that $N_{22}\to 1/A$ which is the behavior required for $N_{22}=t_{22} D_{22}/A^2$, taking into account the  high-energy behavior of $D_{22}(A)$ just discussed.}  The price to pay for having included the second subtraction is the need for an input value for $k_d^ 2$, which has to be provided. It is then more natural for the $^3S_1\text{--}^3D_1$ system to fix the binding energy of the deuteron to its experimental value than the scattering length, as we do below.

As stated in Sec.~\ref{sec:formalism}, an iterative procedure is followed in order to obtain our final results for the phase shifts and the mixing angle from the three $N/D$ equations coupled. For every iteration along that procedure, one obtains  from the $^3 S_1$ wave amplitude the deuteron pole position, $k_d^2$. This is the value used as an input for the function $D_{22}(A)$ at every step. In this way it is not fitted as a free parameter in order to fix the deuteron binding energy, but it comes out in a natural way from $^3 S_1$ and the coupled-channel mechanism. The results that we obtain with this approach are shown in Fig.~\ref{fig_3s13d1_deuteron} by the dashed (blue) lines, while those obtained when there is no deuteron pole in $^3D_1$, using Eq.~\eqref{dd.final.cou} instead of Eq.~\eqref{gdij} with $\ell_{ij}=2$,  correspond to the solid (black) lines. The results are compared with the Nijmegen PWA \cite{Stoks:1994wp} given by the dash-dotted (red) lines. For the $^3S_1$ phase shifts both lines are very similar. The differences are larger for the $^3D_1$ phase shifts, which are then quite sensitive to reproducing correctly the deuteron pole also in the $^3D_1$ partial wave. Indeed, the result without imposing the deuteron in this partial wave is very similar to that obtained from perturbative OPE \cite{peripheral}. Differences are rather small for the mixing angle $\epsilon_1$. As the main contribution to the deuteron comes from $^3 S_1$, its position remains almost unchanged compared with the uncoupled case, with a value obtained for the binding  $E_d\simeq 1.7\ \text{MeV}$, once the experimental scattering length is fixed. This corresponds to an effective range  $r\simeq 0.46$~fm, which is much smaller than the experimental value $r=1.749$~fm, the difference being around a 70\%. This fact is already well documented in the literature \cite{mandelstam}. Indeed, Ref.~\cite{wong1} shows that when the $N/D$ method is used with only OPE as the source of the imaginary part along the LHC, one needs to fit two experimental inputs for every $NN$ $S$ wave in order to reproduce the scattering length and effective range. For $^3S_1$ the scattering length and the deuteron binding energy are taken (we take the same input in Sec.~\ref{oes} below), while for $^1S_0$ two well measured phase shifts at different energies are employed. This result from Ref.~\cite{wong1}, and our own ones presented below in Sec.~\ref{oes}, makes us confidence that a NLO study in ChPT with the $N/D$ method will be phenomenologically successful because a new counterterm enters at this order, multiplying an energy dependent monomial. The authors of Ref.~\cite{wong1} make the approximation of considering only elastic unitarity for $^3S_1$, neglecting its coupling with $^3D_1$, while our treatment is exact.
  
It is also interesting to fix the subtraction constant $N_0$ in terms of the deuteron binding energy and then compare with our previous results when the scattering length was fixed. Imposing $D_{11}(k_d^2)=0$ from Eq.~\eqref{dd.final.3s1} and solving for $N_0$, one has
\begin{align}
N_0&=
\frac{\displaystyle 1+\frac{k_d^2}{\pi}\int_{-\infty}^L \rsps
dk^2 \Delta_{11}(k^2)d_{11}(k^2)g_{11}(k^2,k_d^2)/k^2}{k_d^2\left(g_{11}(k_d^2,0)+{\cal G}(k_d^2)\right)}~,
\label{n0.result}
\end{align} 
where we have first split
\begin{align}
D_{11}(A)=d_{11}(A)-k^2 N_0 g_{11}(A,0)~,
\label{d11.split}
\end{align}
from where the function $d_{11}(A)$ is defined. We have also introduced in Eq.~\eqref{n0.result} the function ${\cal G}(A)$ given by
\begin{align}
{\cal G}(A)=\frac{1}{\pi}\int_{-\infty}^L \rsps dk^2 \Delta_{11}(k^2) g_{11}(k^2,0) g_{11}(A,k^2)~.
\end{align}
The integral equation for $d_{11}(A)$ can be obtained from that in Eq.~\eqref{dd.final.3s1} taking into account Eq.~\eqref{d11.split} and replacing $N_0$ with its expression Eq.~\eqref{n0.result}. This results in
\begin{align}
d_{11}(A)&=1
+\frac{A}{\pi}\int_{-\infty}^L \rsps dk^2 \frac{\Delta_{11}(k^2)d_{11}(k^2)}{k^2}
\left\{
g_{11}(A,k^2) \vphantom{\frac{g_{11}(k^2,k_d^2) {\cal G}(A)}{g_{11}(k_d^2,0)+{\cal G}(k_d^2)}} \right. \\ & \left. -\frac{g_{11}(k^2,k_d^2) {\cal G}(A)}{g_{11}(k_d^2,0)+{\cal G}(k_d^2)}
\right\}
-\frac{A}{k_d^2}\frac{{\cal G}(A)}{g_{11}(k_d^2,0)+{\cal G}(k_d^2)}~. \nn
\end{align}

As in the previous case we fix the dependence on the input by requiring that the deuteron pole  in the mixing wave is located at the same position as in the  $^3S_1$ wave, at the $k_d^2$ corresponding to the binding energy $E_d=2.2~$MeV. Regarding the $^3D_1$ partial wave no pole position is found unless one imposes it in the $D_{22}(A)$ function, making use of Eq.~\eqref{gdij}, having then the right pole structure. Once the deuteron pole is imposed the value that we obtain for the $^3S_1$ scattering length is 4.6~fm and for the effective range 0.41~fm. The latter is indeed very similar to the values obtained before when the scattering length was taken as input. The resulting scattering length is about 15\% lower than its experimental value. We show in Fig.~\ref{fig_3s13d1_deuteron} by the dash-double-dotted (cyan) lines the results obtained when the deuteron pole is imposed in the $^3S_1$ and  $^3D_1$ partial waves, while the double-dotted (green) line is for the case when the deuteron pole position is imposed only in the former.  The results are rather similar to the case when the scattering length was fixed. The most sensitive observable is the mixing angle $\epsilon_1$ where the largest difference happens in the   peak, somewhat less than $1^\circ$. 

\begin{figure}[b]\centering
\includegraphics[width=5.5cm]{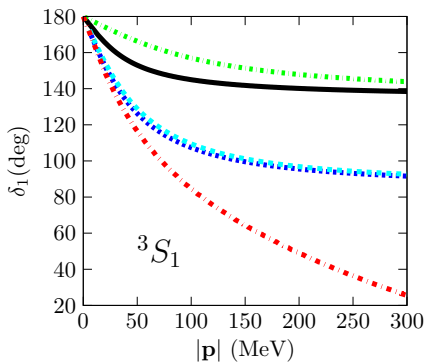} 
\caption{(Color online) Comparison between the results obtained for the theory without pions and our full results at LO. The  Nijmegen PWA \cite{Stoks:1994wp} corresponds to the dot-dashed (red) lines. When the  $^3S_1$ scattering length is fixed one has the dashed (blue) line for the pionfull case and the solid (black) line for the pionless one. When the deuteron binding energy is fixed  the dash-double-dotted (cyan) line for the pionfull theory and the double-dotted (green) line for the pionless case result. 
\label{fig_3s13d1_cc}}
\end{figure}

It is worth comparing our results with the pionless effective field theory. In this case pions are integrated out as heavy degrees of freedom. We can reach this limit by taking $g_A\to 0$ in our results, which implies $\Delta_{ij}=0$. Only the term proportional to $N_0$ survives in  Eq.~\eqref{dd.final.3s1} and $N_{11}(A)=N_0$ from Eq.~\eqref{nn.cou.3s1}. We can determine $N_0$ by fixing the experimental scattering length, Eq.~\eqref{n0.fix}, or by reproducing the deuteron binding energy $N_0=-4\pi/\sqrt{m^3 E_d}$. The former case is given by the solid (black) line and the latter by the double-dotted (green) one in Fig.~\ref{fig_3s13d1_cc}. For comparison we also show the lines corresponding to our full results, obtained by fixing the scattering length and the deuteron binding energy to their experimental values. The former case corresponds to the dashed (blue) line and the latter to the dash-double-dotted (cyan) line, as shown in Fig.~\ref{fig_3s13d1_deuteron}. One observes that the inclusion of pions significantly improves the phase shifts and also makes the results more stable independently of whether the scattering length or the deuteron pole are adjusted.

\begin{figure*}[h!t]
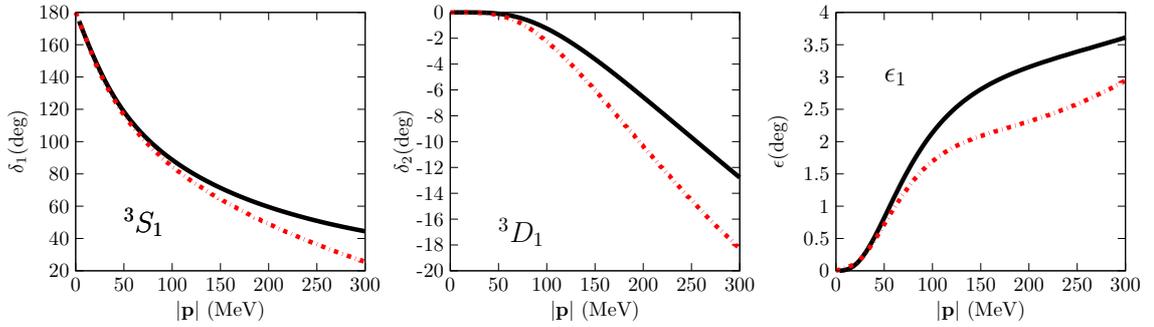
\centering
\begin{tabular}{ccc}
\includegraphics[height=4.2cm,keepaspectratio]{figurasNN_33.eps} &
\includegraphics[height=4.2cm,keepaspectratio]{figurasNN_34.eps} &
\includegraphics[height=4.2cm,keepaspectratio]{figurasNN_35.eps}
\end{tabular}
\caption{(Color online) From left to right we plot our results,  solid (black) curves, for the $^3 S_1$ and $^3 D_1$ phase shifts and the mixing angle $\epsilon_1$, when the experimental $^3S_1$ scattering length and deuteron binding energy are imposed.  The Nijmegen PWA \cite{Stoks:1994wp} data are shown by the dot-dashed (red) lines. 
\label{fig_3s13d1_deuteron_at}}
\end{figure*}

\subsection{One extra subtraction}
\label{oes}

Now we impose that the $^3S_1$ partial wave reproduces the experimental values for the $^3S_1$ scattering length, $a_t$, and the deuteron binding energy simultaneously. Similar restrictions were already considered in Refs.~\cite{wong1,oteo:1989}. To accomplish it we introduce one extra subtraction constant in the $D_{11}(A)$ function by taking one more subtraction in the DR. In this way we enhance the role played by the low-energy region because the extra subtraction gives more weight to the low-energy part of the integrand in the DR, so that it vanishes more rapidly as $A\to \infty$. The new DRs for $N_{11}(A)$ and $D_{11}(A)$ read
\begin{align}
\label{d11.exsb}
N_{11}(A)&=N_0+\frac{A}{\pi}\int_{-\infty}^L dk^2 \frac{\Delta_{11}(k^2)D_{11}(k^2)}{k^2(k^2-A)}~,\nn\\
D_{11}(A)&=1-\frac{A}{k_d^2}-\frac{A(A-k_d^2)}{\pi}N_0\int_0^{\infty}dk^2 \frac{\nu_{11}(k^2)}{(k^2-A)(k^2-k_d^2)k^2}\nn\\
&+\frac{A(A-k_d^2)}{\pi}\int_{-\infty}^Ldk^2 \frac{\Delta_{11}(k^2)D_{11}(k^2)}{k^2} g_{11}^{(d)}(A,k^2) ~,
\end{align}
with
\begin{align}
g_{11}^{(d)}(A,k^2)=\frac{1}{\pi}\int_0^\infty dq^2\frac{\nu_{11}(q^2)}{(q^2-k^2)(q^2-A)(q^2-k_d^2)}~.
\end{align}
By construction $D_{11}(k_d^2)=0$  in Eq.~\eqref{d11.exsb}, which guarantees the presence of the deuteron in its experimental position. Having the right value for the $^3S_1$ scattering length fixes the constant $N_0$ to Eq.~\eqref{n0.fix}. The extra subtraction taken in $D_{11}$, Eq.~\eqref{d11.exsb}, will also be studied when considering the NLO ChPT contribution to the discontinuity across the LHC because then the resulting $\Delta_{ij}(A)$ diverges as $A$ for $A\to\infty$.

The deuteron pole is also imposed in the $^3D_1$ partial wave by employing Eq.~\eqref{gdij} so that the right pole structure is accomplished. The input is fixed such that the resulting deuteron pole position in 
the mixing partial wave $^3S_1$-$^3D_1$ is located in the same position as for the other two coupled partial waves, as already discussed above.

In Fig.~\ref{fig_3s13d1_deuteron_at} we show, from left to right, the $^3S_1$ and $^3D_1$ phase shifts and  the mixing angle $\epsilon_1$ resulting from Eq.~\eqref{d11.exsb}, in that order. A clear improvement as compared with Fig.~\ref{fig_3s13d1_deuteron} is observed, so that now the resulting curve run closer to the Nijmegen PWA \cite{Stoks:1994wp} for the $^3S_1$ phase shifts. An improvement also happens for the mixing angle $\epsilon_1$ which now overlaps better with the Nijmegen results for three-momentum up to about $100 \text{\ MeV}$ and later the trend of the curve tends to follow that of the Nijmegen PWA. Let us also stress that the failure to reproduce $\epsilon_1$ in the Kaplan-Savage-Wise scheme \cite{kswa} was the main reason to conclude that its perturbative treatment of pion exchange was not appropriate \cite{fleming:2000}. In contrast, our LO reproduction of $\epsilon_1$ in Fig.~\eqref{fig_3s13d1_deuteron} is already quite close to the Nijmegen results \cite{Stoks:1994wp} and improves when considering the extra subtraction, as shown in Fig.~\ref{fig_3s13d1_deuteron_at}. This is a clear indication that $\epsilon_1$ will be also properly reproduced at NLO in the calculation of $\Delta_{ij}(A)$, although the adjusted value for this subtraction constant would change due to the addition of the two-pion exchange contributions.

Next we evaluate the three independent deuteron parameters that can be calculated from $NN$ scattering \cite{swart1}. The first quantity is the binding energy of the deuteron that is fixed to its experimental value as input. The second quantity that we consider is the asymptotic $D/S$ ratio $\eta$. For that we make use of the Blatt and Beidenharn parameterization \cite{blatt} and diagonalize the $^3S_1$-$^3D_1$ $S$-matrix, $S_{1}$, by an orthogonal real matrix ${\cal O}$,
\begin{align}
S_{1}&={\cal O}\,S_{1;\rm{diag}}\,{\cal O}^{-1}~,\nn\\
{\cal O}&=\left(\begin{array}{ll}
\cos \epsilon_1 & -\sin\epsilon_1\\
\sin \epsilon_1 & \cos\epsilon_1
\end{array}
\right)~,\nn\\
S_{1;\rm{diag}}& \equiv \left(
\begin{array}{ll}
S_0 & 0\\
0 & S_2
\end{array}
\right)~.
\end{align}
In terms of $\epsilon_1$ one can write for the asymptotic $D/S$ ratio $\eta$ as \cite{swart1,swart2}
\begin{align}
\eta=-\tan\epsilon_1~.
\end{align}
The third quantity that we calculate is $i$ times the residue of the eigenvalue $S_0$ at the deuteron pole position $\alpha\equiv \sqrt{-k_d^2}$,
\begin{align}
S_0=\frac{N_p^2}{\alpha+i |\vp|}+\rm{{\small regular~ terms}}~.
\end{align}

We should remark that because we do not employ $NN$ potential to study $NN$ scattering we cannot compute the wave function of the deuteron and in terms of it evaluate straightforwardly (in the simplest approximation) other quantities, e.g., the deuteron electric quadrupole moment $Q$ or the mean-square deuteron radius $\la r^2\ra^{1/2}$. This does not mean that we cannot obtain such observable quantities from our $T$-matrix but simply  that we should consider other processes beyond pure $NN$ scattering. For instance, in order to calculate the mean-square deuteron radius $\la r^2\ra $ we should proceed as in Ref.~\cite{sigalba} to calculate the same quantity but for the $f_0(500)$ or $\sigma$ 
resonance, where $\pi\pi$ scattering in the presence of a scalar source was calculated. Similarly, we should study here $NN$ scattering in the presence of a scalar source giving rise to the matter form factor of the deuteron. This is beyond the present study and requires an independent study. 

The resulting values that we obtain are 
\begin{equation}
\eta  = 0.028~, \quad
N_p^2 = 0.74~\text{fm}^{-1}.
\label{paramdeu}
\end{equation}
 Our results 
 compare well with the experimental determinations, $\eta=0.0271(4)$ \cite{9.10.fri} and
 $\eta=0.0263(13)$ \cite{9.klars}. They are also close to those evaluated in Nijmegen PWA 1993 \cite{Stoks:1994wp} 
\begin{equation}
\label{eta}
\eta  = 0.02543(7)~, \quad
N_p^2 =0.7830(7)~\rm{fm}^{-1}~.
\end{equation}

Thus, once we reproduce simultaneously the deuteron binding energy and the $^3S_1$ scattering length, the other properties of the deuteron that can be extracted from scattering  compare well with the values determined in partial-wave analyses or experiment.   

We obtain the following value for the effective range $r$, 
\begin{align}
\label{rd}
r&=1.56(3)~\text{fm}~,
\end{align}
where the error is just statistical by fitting the low-energy phase shifts generated by 
our own amplitudes. This number is 
quite close to the Nijmegen PWA 1993 \cite{Stoks:1994wp} result, $r=1.753(2)$~fm. 

In Ref.~\cite{frederico:1999} the OPE potential from ChPT is employed in a LS equation solved by 
making use of an interesting method based on identifying the input with the $T$-matrix deep in the LHC, writing the potential in terms of it. Their results for $\eta$ and $r$ are very similar to ours  in Eqs.~\eqref{eta} and \eqref{rd}, obtaining  the intervals of values $\eta=0.0281\text{--}0.0293$ and $r=1.36\text{--}1.58$~fm. Their results for the elastic $^3S_1$  phase shifts are also quite similar to ours, though for $^3D_1$ they are closer to  Nijmegen points \cite{Stoks:1994wp}. Regarding the mixing angle $\epsilon_1$, Ref.~ \cite{frederico:1999} obtains that for a large renormalization scale $\mu$  the resulting curves  depart from the Nijmegen data \cite{Stoks:1994wp} by an absolute amount similar to ours  for $|\vp|\gtrsim 100$~MeV (our results lie above, while theirs lie below). One should keep in mind that we have taken the  scattering length and the binding energy as input for our calculations, while Ref.~\cite{frederico:1999} only adjusts the scattering length. 

It is well known since the 1960s that for $^3S_1\text{--}^3D_1$ coupled partial waves, solving a LS equation in terms of the OPE potential gives a significantly better phenomenology than solving the $N/D$ method taking for $\Delta_{ij}(A)$ the discontinuity along the LHC induced by OPE \cite{noyes65}. However, it is worth keeping in mind that  \cite{frederico:1999}, as well as \cite{nogga}, obtain phase shifts for  $^1S_0$ which are very similar to ours in Ref.~\cite{paper1}. It is known that the $^1S_0$ phase shift data of Nijmegen \cite{Stoks:1994wp} are reproduced quite closely \cite{epen3lo} once two-pion exchange contributions and NLO LECs in the four-nucleon Lagrangian are included. In our novel theory, which calculates the $NN$ partial waves from  ChPT by employing the $N/D$ method, there is no reason to expect that the phase shifts should be reproduced at LO worse in  the $^1S_0$ partial wave than in the $^3S_1\text{--}^3 D_1$  coupled waves (which results strictly correspond to  Fig.~\ref{fig_3s13d1_deuteron} in terms of only one subtraction being needed).  In this respect, it is rewarding that, by considering NLO contributions to the $NN$ potential in the standard Weinberg approach \cite{epen3lo}, one can obtain good results for $^1S_0$. This should be also expected for the $^3S_1\text{--}^3D_1$ case within our approach. Indeed, we have already seen that, by including one extra subtraction, the reproduction of phase shifts (particularly for $^3S_1$) and mixing angle clearly improves. When considering two-pion exchange at NLO some extra counterterms are needed because $\Delta_{ij}(A)$ diverges as $A$ for $A\to -\infty$ along the LHC.

Solving a LS equation with OPE for the $^3S_1\text{--}^3D_1$ system is much more successful phenomenologically than for the $^1S_0$ case. One should be aware that this is something that is checked {\it a posteriori} and is not rooted in the chiral counting (in which our approach is based). From our point of view the ladder resummation in the LS for the $^3S_1\text{--}^3D_1$ case provides higher orders terms to $\Delta_{ij}(A)$ in the right direction. However, this improvement should come out when applying the $N/D$ method to ({\it just } a few) higher orders, because along the LHC $\Delta_{ij}(A)$ is perturbative and amenable to a chiral expansion as discussed. For $^1S_0$ the higher orders in $\Delta(A)$ provided by the LS equation are not the important source of dynamics and one has to really consider the full machinery in order to incorporate at higher orders two-pion exchange with the associated chiral counterterms. It is our aim to develop for the time being a NLO (or NNLO) study of $NN$ scattering with our approach based on the $N/D$ method and the ChPT calculation of 
$\Delta(A)$ in order to definitively settle this important issue. We would like to stress that at this stage our study is mostly exploratory and not competitive with the current sophisticated potentials \cite{Stoks:1994wp} or calculated at higher orders from ChPT \cite{entem,epen3lo}.

 The set of works \cite{friar:1984,ballot:1992,sprung:1994,pavones,pavones2} gives rise to a remarkable description of deuteron properties employing the $NN$ potential given by OPE in a LS equation, e.g.,  Ref.~\cite{pavones}  achieves for many observables  a $2-3\%$ of  deviation with respect to the experimental values. But this is not the only aim of an EFT. That is, one does not expect such a high degree of convergence by taking only the LO ChPT $NN$ potential. This is more a matter of  phenomenological success and not rooted in the chiral EFT. For baryon ChPT the expansion scale is not so great, $\Lambda \simeq 12 \pi^2 f_\pi^2/g_A^2 m \simeq 500$~MeV \cite{long,talk}, and such a great precision is thus difficult to understand from the ChPT expansion. We want to emphasize this point (consider, e.g., the not so great achievement for the $^1S_0$ case)  and develop a formalism where contributions to a given process can be obtained order by order systematically in the chiral EFT expansion of $\Delta(A)$.

\subsection{Higher partial waves}
\label{sec:results}


\begin{figure}[t]
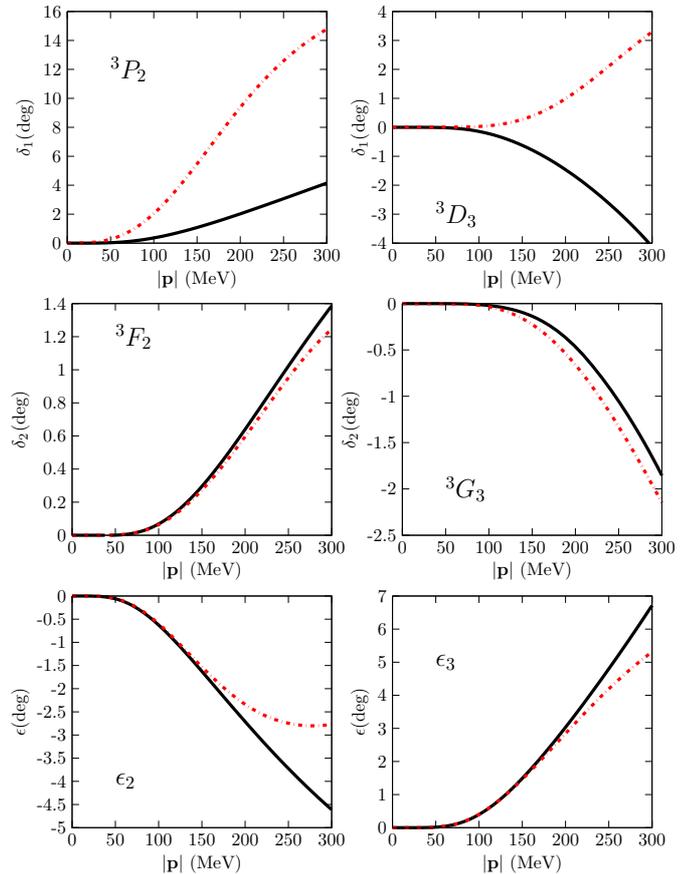

\centering
\begin{tabular}{cc}
\includegraphics[height=0.21\textwidth,keepaspectratio,trim = 2mm 0mm 2mm 0mm]{figurasNN_3.eps} &
\includegraphics[height=0.21\textwidth,keepaspectratio,trim = 2mm 0mm 2mm 0mm]{figurasNN_6.eps} \\
\includegraphics[height=0.21\textwidth,keepaspectratio,trim = 2mm 0mm 2mm 0mm]{figurasNN_4.eps} &
\includegraphics[height=0.21\textwidth,keepaspectratio,trim = 2mm 0mm 2mm 0mm]{figurasNN_7.eps} \\
\includegraphics[height=0.21\textwidth,keepaspectratio,trim = 2mm 0mm 2mm 0mm]{figurasNN_5.eps} &
\includegraphics[height=0.21\textwidth,keepaspectratio,trim = 2mm 0mm 2mm 0mm]{figurasNN_8.eps}
\end{tabular}
\caption{(Color online) Comparison of our results for the $^3 P_2$, $^3 F_2$, $^3 D_3$ and $^3 G_3$ phase shifts and the mixing angles $\epsilon_2$ and $\epsilon_3$, shown by the solid (black) lines, to the Nijmegen PWA \cite{Stoks:1994wp}, represented by the dot-dashed  (red) lines.\label{fig_3p23f2_3d33g3}}
\end{figure}

In this section, we present the results for the spin triplet waves with total angular momentum $J=2\text{ and }3$, obtained with the formalism derived in Sec.~\ref{sec:formalism}. They are shown by the solid (black) lines in Fig.~\ref{fig_3p23f2_3d33g3}, where they are compared with the Nijmegen PWA \cite{Stoks:1994wp} [dash-dotted (red) lines].

We already see a good agreement with data for $^3F_2$ and $^3G_3$ as well as for the mixing angles $\epsilon_2$ and $\epsilon_3$. The lower partial waves $^3P_2$ and $^3D_3$ are not well reproduced with only OPE yet. This fact for the $^3D_3$ partial wave was already observed in Ref.~\cite{peripheral}, where OPE was treated perturbatively. In this reference $^3D_3$ is also obtained with opposite sign to the data. In Ref.~\cite{nogga}, with one counterterm promoted to LO for the $^3P_2$ wave, the situation is similar. The $^3P_2$ and $^3D_3$ phase shifts are not well reproduced at LO, while the others compare well with the data. We expect to restore the agreement with experiment at higher orders in the application of our method to $^3P_2$ and $^3D_3$.

\section{Conclusions}\label{sec:conclusions}
We have developed a new set of equations for the $N/D$ method in coupled partial waves, extending our previous work, Ref.~\cite{paper1}, restricted to uncoupled partial waves. This method is presented in a novel way, adequate to improve the results systematically by taking higher orders in the chiral expansion of the calculation of the discontinuity of the partial wave amplitudes along the LHC, $\Delta_{ij}(A)$ with $A< 4m_\pi^2$. This extension is accomplished by providing three $N/D$ equations for each set of partial waves coupled. The solution is obtained in an iterative and self-consistent way. The correct solution satisfies unitarity and for the case of the $^3S_1\text{--}^3D_1$ system the deuteron pole is located at the same position in all the waves, having the correct pole structure.
	
As in Ref.~\cite{paper1} our approach guarantees the right threshold behavior for partial waves with orbital angular momentum $\ell_{ij} \geq 2$ by satisfying  $\ell_{ij}-1$ constraints. Since the function $D_{ij}(A)$ is determined modulo the addition of Castillejo-Dalitz-Dyson poles (that correspond to  zeros of the $NN$ partial waves along the real axis) we have then added $\ell_{ij}-1$ of such poles at infinity in $D_{ij}$   for $\ell_{ij}\geq 2$. By sending such poles to infinity we do not include any zero of any $NN$ partial wave at finite energies. In addition, the residues of these poles in $D_{ij}(A)$ are fixed once the sum rules are satisfied, so that no new parameters are included. At low energies the CDD poles behave like adding a polynomial of degree $\ell_{ij}-2$ to $D_{ij}(A)$. 

We have studied the $^3S_1\text{--}^3D_1$ coupled waves by fixing the resulting subtraction constant to either the experimental value of the $^3S_1$ scattering length or the deuteron binding energy. We find that the $^3D_1$ phase shifts are the most sensitive to this choice. As expected, in all cases the triplet $S$ wave effective range comes out much smaller than in experiments. Then, we added one extra subtraction to calculate the $^3S_1$ wave requiring the simultaneous reproduction of the deuteron binding energy and triplet $S$ wave scattering length. The resulting $^3S_1$ phase shifts are much improved and we then obtain the effective range and deuteron properties close to their experimental values.  We have also considered the pionless case and compared with our full results, which include OPE. It was seen then that the results clearly improve for the latter case. For the waves with orbital angular momentum  $\ell\geq 1$ at LO there is no subtraction constant and the results are parameter free. The resulting phase shifts and mixing angles agree well with the Nijmegen PWA results, except for the $^3P_2$ and $^3D_3$ partial waves. 

Certainly, including OPE as the only source of discontinuity along the LHC is phenomenologically just a first step and a NLO calculation should be undergone to establish the capability of the method to reproduce properly $NN$ scattering data. However, one should stress at this point that our approach based on the $N/D$ method offers a way to calculate $NN$ scattering independently of  cutoff, because only convergent integrals appear, while keeping the chiral power counting. The dispersive integrals are convergent  by taking the appropriate number of subtractions with the related subtraction constants fixed to experimental data. At LO only two subtraction constants appear in the $^1S_0$ and $^3S_1$ partial waves, the same number as LO ChPT counterterms \cite{weinn}. This method allows one to perform calculations systematically, order by order, in ChPT.

\begin{acknowledgments}
 We  would like to thank Ulf-G.~Mei{\ss}ner for valuable discussions and a critical reading of the manuscript. His warm hospitality at the Helmholtz Institut f\"ur Strahlen- und Kernphysik, where part of this work was performed, is also acknowledged. This work is partially funded by MICINN Grant No.~FPA2010-17806 and Fundaci\'on S\'eneca Grant No.~11871/PI/09. We also appreciate the financial support from  the BMBF Grant No.~06BN411, the EU-Research Infrastructure Integrating Activity ``Study of Strongly Interacting Matter'' (HadronPhysics2, Grant No. 227431) under the Seventh Framework Program of EU and the Consolider-Ingenio 2010 Programme CPAN (CSD2007-00042).
\end{acknowledgments}

\appendix*
\section{The \boldmath $g_{ij}(A,k^2)$ functions}\label{app:gfun_div}
On general  grounds, the following threshold behavior are found for the $\nu_{ij}(A)$ functions:
\begin{equation*}
\nu_{11}(A) \propto A^{1/2} \quad \nu_{12}(A) \propto A^{-1/2} \quad \nu_{22}(A) \propto A^{-3/2}
\end{equation*}
This can also be seen by inserting the low energy behavior of $\delta_1$ ($\propto A^{\ell+1/2}$), $\delta_2$ ($\propto A^{\ell' + 1/2}$), and $\epsilon$ ($\propto A^{(\ell+\ell'+1)/2}$) in the explicit expression for $\nu_{ij}(A)$, Eqs.~\eqref{eq:nus11}--\eqref{eq:nus12}. No problem occurs in the integrand for the functions $g_{11}(A,k^2)$ and $g_{12}(A,k^2)$, when these low-energy behaviors are inserted, but the divergence in $\nu_{22}(A)$ could lead to a divergence in the function $g_{22}(A,k^2)$. This was already pointed out in Ref.~\cite{noyes}, as a potential source of divergences. However, a more careful analysis shows that this divergence vanishes owing to the sum rules, Eqs.~\eqref{n.sr.cou}. For the $g_{22}(A,k^2)$ integral one has
\begin{align*}
g_{22}(A,k^2) = & \frac{1}{\pi}\int_{\lambda \to 0}^{+\infty} \rspo \dd q^2 \frac{\nu_{22}(q^2)}{(q^2-A)(q^2-k^2)} \\
= & \frac{2\nu_0}{\pi Ak^2 \sqrt{\lambda}} + \text{regular terms}
\label{int.g22}
\end{align*}
where $\nu_{22}(A) = \nu_0 A^{-3/2}$ for $A \to 0$. In the previous equation the regular terms refer to the rest of the contributions to the integral, which do not diverge for $\lambda\to 0$.  The divergent term in the previous equation  enters into the integral  Eq.~\eqref{dd.final.cou} through the function $g_{22}(A,k^2)$, giving rise to a term proportional to
\begin{equation*}
\int_{-\infty}^L \rsps \dd k^2 \frac{\Delta_{22}(k^2)D_{22}(k^2)}{k^4}=0,
\end{equation*}
which vanishes owing to the constraints of Eq.~\eqref{n.sr.cou}. Note that every channel in which  $g_{22}(A)$  is involved  has $\ell' \geq 2$ (the lowest value for $\ell'$ corresponds to the $^3D_1$ wave), and thus the  sum rule above applies. The constraints  Eq.~\eqref{n.sr.cou} thus show a new important facet  beyond the original motivation for their introduction. 

\bibliography{apsrev}

\end{document}